**Developing a Machine-Learning Algorithm to Diagnose Age-Related Macular Degeneration**

Ananya Dua* (1), Pham Hung Minh* (2), Sajid Fahmid* (3), Shikhar Gupta*  (1), Sophia Zheng* (4), Vanessa Moyo* (5), Yanran Elisa Xue* (6)

1. Monta Vista High School
2. Hanoi-Amsterdam High School
3. Churchill Fulshear High School
4. Norfolk Academy
5. John H. Guyer High School
6. Collège Sainte-Anne

* These authors contributed equally.

**ABSTRACT**

Today, more than 12 million people over the age of 40 suffer from ocular diseases. Most commonly, older patients are susceptible to age related macular degeneration, an eye disease that causes blurring of the central vision due to the deterioration of the retina. The former can only be detected through complex and expensive imaging software, markedly a visual field test; this leaves a significant population with untreated eye disease and holds them at risk for complete vision loss. The use of machine learning algorithms has been proposed for treating eye disease. However, the development of these models is limited by a lack of understanding regarding appropriate model and training parameters to maximize model performance. In our study, we address these points by generating 6 models, each with a learning rate of $1 * 10^n$ where n is 0, -1, -2, ... -6, and calculated a f1 score for each of the models. Our analysis shows that sample imbalance is a key challenge in training of machine learning models and can result in deceptive improvements in training cost which does not translate to true improvements in model predictive performance.

Considering the wide ranging impact of the disease and its adverse effects, we developed a machine learning algorithm to treat the same. We trained our model on varying eye disease datasets consisting of over 5000 patients, and the pictures of their infected eyes. In the future, we hope this model is used extensively, especially in areas that are under-resourced, to better diagnose eye disease and improve well being for humanity.

**INTRODUCTION**

Age related macular degeneration is an ocular disease that affects the central retina at the back of the eye. The disease most commonly leads to blindness, and it involves a gradual deterioration of the retina cells or a fluid buildup in the retina that progressively hinders the sight. Ocular diseases including age-related macular degeneration are found in more than 196 million people aged 40 and up world wide, and these diseases are expected to grow to 240 million by 2050. The above is characterized by abnormalities in vision and retinal pigment. The disease is divided into three distinctions based on the age of the individual because of the prospect of development of advanced AMD. An evaluation for the diagnosis of age-related macular degeneration involves a comprehensive eye exam of the macula that hones in on aspects that are unique qualities of AMD. Factors that may be assessed are decreased vision, difficulty in adjusting to darkness, and a history of ocular issues. The process of gathering this information can be time consuming and requires advanced machinery to produce a clear image of the macula. One example of the type of machinery required to diagnose age-related macular degeneration successfully is an optical coherence tomography (OCT). The issue is not only on the premise of efficiency but costs too. The cost of purchasing an OCT can range from $40,000 to $150,000. This cost can be a heavy burden for small eye clinics that are sporadically found in the regions outside of a city. As a result, we can find these advanced types of diagnostic machinery in inner city large hospitals and large eye clinics. Access to these types of machinery is a challenge that with affordable and efficient options will drastically lower the inequities regarding access to eye treatment.

As a result of the issues regarding the diagnostic processes of age-related macular degeneration a type of technology known as a convolutional neural network (CNN), a deep learning algorithm designed specially for images and pixel processing, can be especially used effectively in this scenario. This network is made up of three layers: a convolutional layer, a pooling layer, and finally the fully connected/output layer. They work in unison to take in image inputs and output a 4D array, which houses predictions and conclusions regarding the data. A CNN completes this task in a fraction of the time it would take with conventional methods, and it is functional in use for other contexts as well, especially in areas where a comprehensive exam may not be feasible. CNN models can be trained up to 99% accuracy; ours reached ___. This process has high yielding results; strategies that are needed to train CNN models for the assessment and diagnosis of ocular diseases have not been well characterized. By these ends, in this study, we aimed to train the model on 5000 images of eyes affected by age-related macular degeneration, to conclude diagnostic accuracy.

**MATERIALS AND METHODS**

Datasets

The source where our dataset is originally from is https://odir2019.grand-challenge.org. The health-related images in the dataset are anonymised and de-identified. The dataset is empirical patient's health information gathered by Shanggong Medical Technology Co., Ltd. from different hospitals/medical centers in China. The data of the patient's' health is categorized and labeled as follows:
- Normal (N),
- Diabetes (D),
- Glaucoma (G),
- Cataract (C),
- Age related Macular Degeneration (A),
- Hypertension (H), Pathological Myopia (M),
- Other diseases/abnormalities (O)

The original dataset is made up of patients whose ages range from 1 to 94, with the majority falling into the 55-64 age group. The distribution of sex is relatively equal, with 54% are male, and 46% are female.

From this dataset, we narrow our research topic down to the "Age-Related Macular Degeneration (A)" and "Normal (N)". There are no missing datas in the final dataset that we use for training our machine learning model.

Convolutional neural network

In our research, we utilize a ResNet 18 convolutional neural network for classification [1].
  A. ReLU Layer:
     - The Rectified Linear Unit layer is ultimately used for enhancing non-linearity in the convolutional neural network

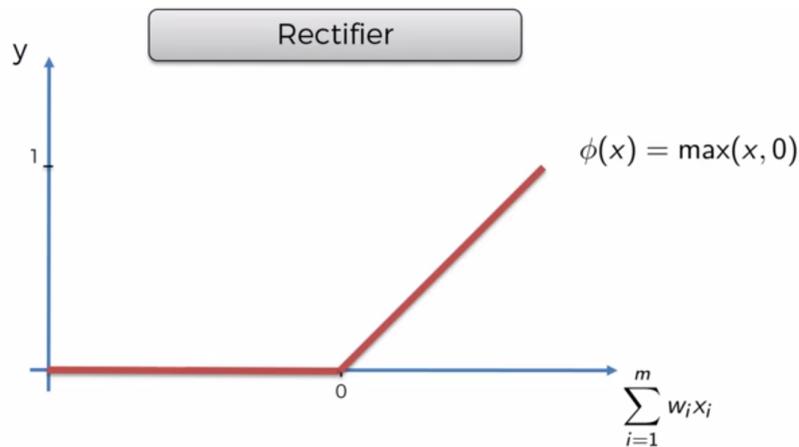
**Figure 1. Title**
What does this figure show

B. Convolution layer :
The basic convolution function is:

$$(f*g)(t) = \int_{-\infty}^{\infty} f(\tau)g(t-\tau)d\tau$$

It is a combined integration of 2 functions that delineates how the shape of one is modified by the other.

The convolution layer consists of kernels, which are going to be trained for optimization. These kernels, whose size is often 3*3, convolve with the input image and generate what are called feature maps. The process continues over the whole dimension of the input image as the kernel strides vertically and horizontally along the input matrix. With different kernels, we manage to generate multiple feature maps by iterating the aforementioned process. These feature maps are stacked among the others, creating a convolutional layer.

C. Pooling layer:
In order to make the convolutional network become more flexible with spatial variance of the inputs, the pooling layer is implemented through leveling down the resolution of the inputs. During the process, the number of parameters is reduced, which also helps prevent overfitting in the model. The pooling layer with pooled layer whose size is 2*2 and stride of two is commonly used, and each pooled feature map matches with the corresponding feature map of the convolutional layer. The popular procedure used in pooling is max pooling and average pooling. For the former method, the max element is picked out from the neighborhood and for the latter, the mean element is selected. The mechanics behind the two types of pooling are as follows:

$$\text{max} : \quad y_{i^{l+1}, j^{l+1}, d} = \max_{0 \leq i < H, 0 \leq j < W} x^l_{i^{l+1} \times H + i, j^{l+1} \times W + j, d},$$

$$\text{average} : \quad y_{i^{l+1}, j^{l+1}, d} = \frac{1}{HW} \sum_{0 \leq i < H, 0 \leq j < W} x^l_{i^{l+1} \times H + i, j^{l+1} \times W + j, d},$$

D. Fully connected layer:
The pooling layer is flattened into a vector, which is used as the input for future fully connected artificial neural networks. The fully connected layer is a type of hidden layer in which each node possesses full connection to all nodes in the next layer. Its aim is to assign input images into specific classes through training.

- Cost function:
- Throughout the process we train the model to minimize the cost function by gradient descent. The optimization concept can be illustrated as:

Gradient descent algorithm helps calculate the minimum

- Performance Metrics
    Fscore
    - Useful for unevenly distributed datasets, and gives a true result that isn't biased to class sizes, and it's really useful in the medical field (takes into account recall and precision, false negatives are really helpful for the medical field).
- We evaluate our model's performance through the F-score. The measurement is calculated through precision and recall, given that they are different from 0:

$$\frac{2}{F1} = \frac{1}{precision} + \frac{1}{recall} \text{ or } F1 = 2\frac{1}{\frac{1}{precision} + \frac{1}{recall}} = 2\frac{precision \times recall}{precision + recall}$$

The F-score value falls into the interval of (0,1]
.

## RESULTS

The main objective of the research was to see if there is a correlation between different learning rates on our pre-compiled convolutional neural network model, Resnet18, offered by the PyTorch library. We ran several tests on different learning rates, and looked at different metrics to gauge the performance of the model on real world cataract imagery.

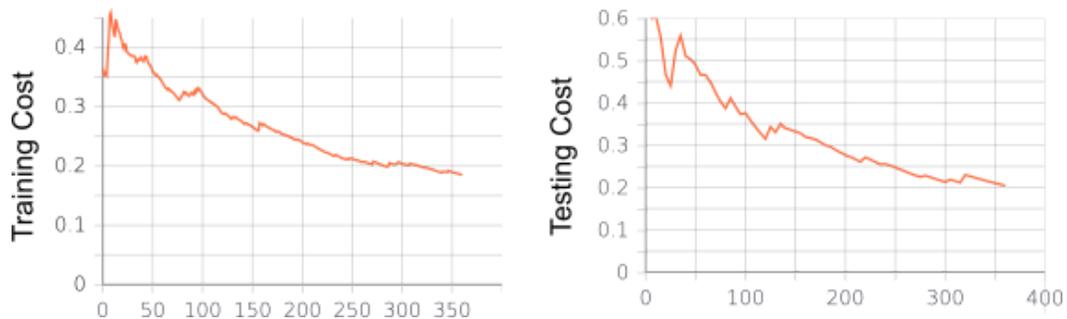

**Figure 1**
Unweighted base model

Initially, the tests were run on unweighted batches of images. An important remark is that our batches were unweighted. Our training set consisted of a significantly greater number of regular, non-cataract diseases images, compared to the images with the cataract. It can be seen that the model has been very well trained, as can be seen by the training and testing costs graphs, which all indicate to be moving down as the model goes through more epochs of training.

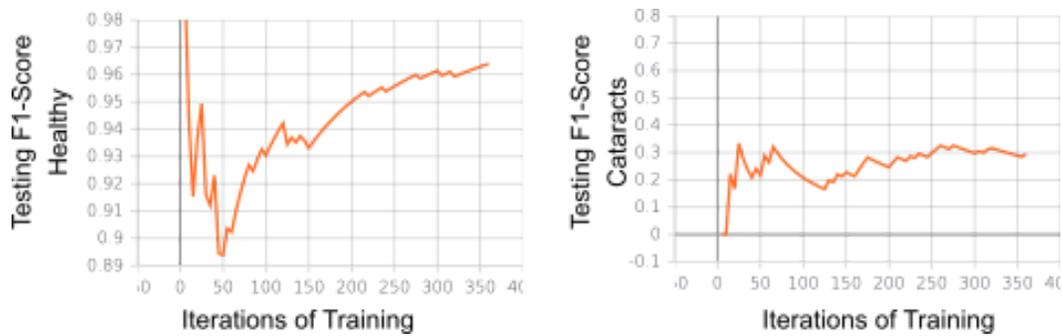

**Figure 2**
Unweighted base model

However, as can be seen by Figure 2, this conclusion by the testing and training loss curve is deceptive. When using the F1 score to assess our model, we saw that the F1 score remained stagnant for the testing set, unlike what we would have assumed when solely looking at the training and testing costs. This is due to the model simply predicting no cataract, and being correct most of the time, because of the imbalance in the dataset classes.

In our next two experiments, we made two main changes: the first one was to weight the batches of the images passed into the neural network, and the second one was to weight the cost function so that it is penalized heavier for misclassifying the cataract image, so that it does not simply predicted no cataract for all images and have a decreasing cost.

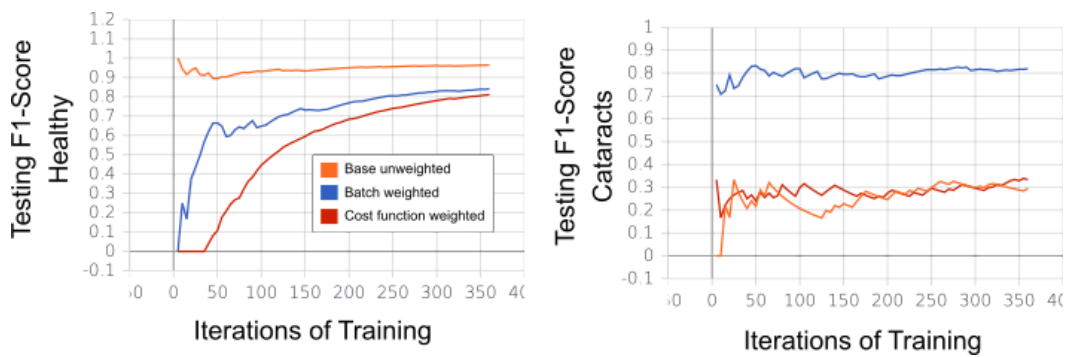

**Figure 3**
Unweighted base model

**DISCUSSION**

- CNN's are really useful for macular degeneration detection, come with f1 of ___%.A
- *Different neural networks/algorithms?*
  Our next target is to make the model as
    - **Higher image size** improves model performance, but there is a compromise on it's learning rate. This would not be applicable in the real world where there would be millions of images.
    - With a higher batch size, there would be images of each category in every batch, so the model wouldn't be predicting the same result every time.
    - Weighting: balances out the difference in number of images in each category

- - The learning rate is a mechanism that fine tunes the model further. We hypothesize that a lower learning rate contributes the best to the goal of lowering the least cost function and reaching the minimum point of the model the quickest. We performed tests using the TensorFlow terminal code in our code and selected 9 learning rates as our sample learning rates to be tested with. After testing the learning rates with the code to test **(or should it be train instead of test?)** we observed that
- Sources of bias in the model
  - **Lower batch size** lowers cost function, but the model isn't actually getting trained because it is simply learning to predict the dominant category –that appears 90% of the time– every time.
    - For example, 1 image per batch: model isn't being trained because it learns by comparing each image to other images in the batch… since there's nothing to compare the image with
  - More 'Normal (N)' images than 'Age related macular degeneration (A)'
- Suggestions for future researchers in this area
  - (what should we put here?) wtv u can think of (yeah I'll try to think of something lol)

## ACKNOWLEDGMENTS

We would like to acknowledge Parsa Akbari for assisting [1] with generation of results and write-up of the report. [1] Downing College, Cambridge University, United Kingdom, CB2 1DQ